\documentclass{PoS}

\title{Application of machine learning techniques at BESIII experiment}

\ShortTitle{Application of machine learning techniques at BESIII experiment}

\author{\speaker{Beijiang Liu}\thanks{This work is supported in part by the CAS Large-Scale Scientific Facility Program; CAS Key Research Program of Frontier Sciences under Contract No. QYZDJ-SSW-SLH040; Joint Large-Scale Scientific Facility Funds of the NSFC and CAS under Contract No. U1732103.}\\
        Institute of High Energy Physics, Chinese Academy of Sciences\\
        E-mail: \email{liubj@ihep.ac.cn}}
\author{Xian Xiong\\
        Institute of High Energy Physics, Chinese Academy of Sciences\\
        E-mail:\email{xiongxa@ihep.ac.cn}}
\author{Guoyi Hou\\
        Institute of High Energy Physics, Chinese Academy of Sciences\\
        }
\author{Shiming Song\\
        Sichuan University\\
        }
\author{Lin Shen\\
        Sichuan University\\
        }

\abstract{BESIII is a currently running tau-charm factory with the largest samples of on threshold charm meson pairs, directly produced charmonia and some other unique datasets at BEPCII collider.
Machine learning techniques have been employed to improve the performance of BESIII software. The studies for reweighing MC, particle identification and cluster reconstruction for the CGEM (Cylindrical Gas Electron Multiplier) inner tracker are presented.}

\FullConference{ICHEP2018, XXXIX International Conference on High Energy Physics\\
		4-11 July 2018\\
		Seoul, Korea}

\begin{document}

\section{Introduction}

The BESIII experiment (Beijing Spectrometer)~\cite{besiii_detector}, located at BEPCII
(Beijing Electron Position Collider), is an experiment at the high precision frontier of hadron physics in $\tau$-charm region. BESIII has been successfully operating since 2008 and has collected the world's largest data samples of $J/\psi$, $\psi(3686)$ and $\psi(3770)$ decays as well as data in the energy region above 4 GeV, which benefit rich physics programs~\cite{phsysics_at_besiii}.
Machine learning (ML) techniques have been employed to improve the performance of BESIII software. Novel approaches for muon identification, multi-dimensional reweighting  and cluster reconstruction for the  Cylindrical Gas Electron Multiplier inner tracker (CGEM-IT) are presented.

\section{A new approach for muon identification}

The BESIII detector has a geometrical acceptance of 93\% of 4$\pi$ and consists a small-celled, helium-based main draft chamber, an electromagnetic calorimeter (EMC), a time-of-flight system (TOF) for particle identification, a muon chamber system (MUC) with resistive plate chambers incorpared in the return iron of the superconducting solenoid with 1T magentic field. More details of the detector are described in Ref~\cite{besiii_detector}.
A combined confidence level for muon identification is formed by using the specific ionization energy loss in the MDC (dE/dx), the time of flight, and the information from EMC and MUC. $\mathcal{L}=\mathcal{L}_{\mathrm{dE/dx}}\cdot\mathcal{L}_{\mathrm{TOF}}\cdot\mathcal{L}_{\mathrm{EMC}}\cdot\mathcal{L}_{\mathrm{MUC}}$, where $\mathcal{L}_{\mathrm{dE/dx(TOF)}}$ is calculated from $\chi^2$ of particle hypothesis ($e, \mu, \pi, K, p$) and $\mathcal{L}_{\mathrm{EMC(MUC)}}$ is a normalized output of a shallow neural network of EMC(MUC). The discriminate of $\mu$ / $\pi$ is critical for many of the analyses. The task is very challenging, because the mass of $\mu$ and $\pi$ are very close and $\mu$ and $\pi$ are difficult to be discriminated with $dE/dx$ and TOF.

\begin{figure}[h]
\centering
\includegraphics[width=5cm,clip]{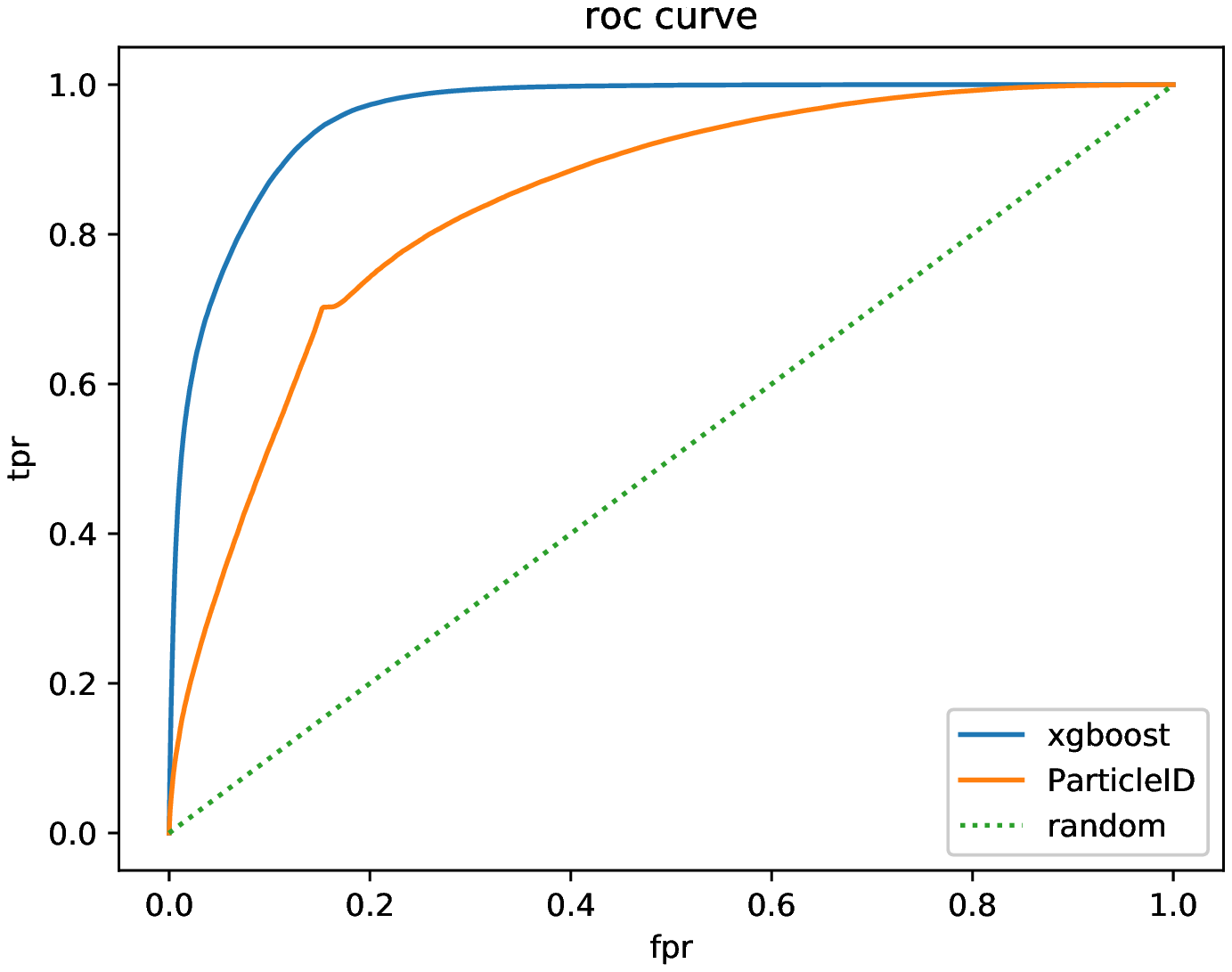}
\includegraphics[width=5cm,clip]{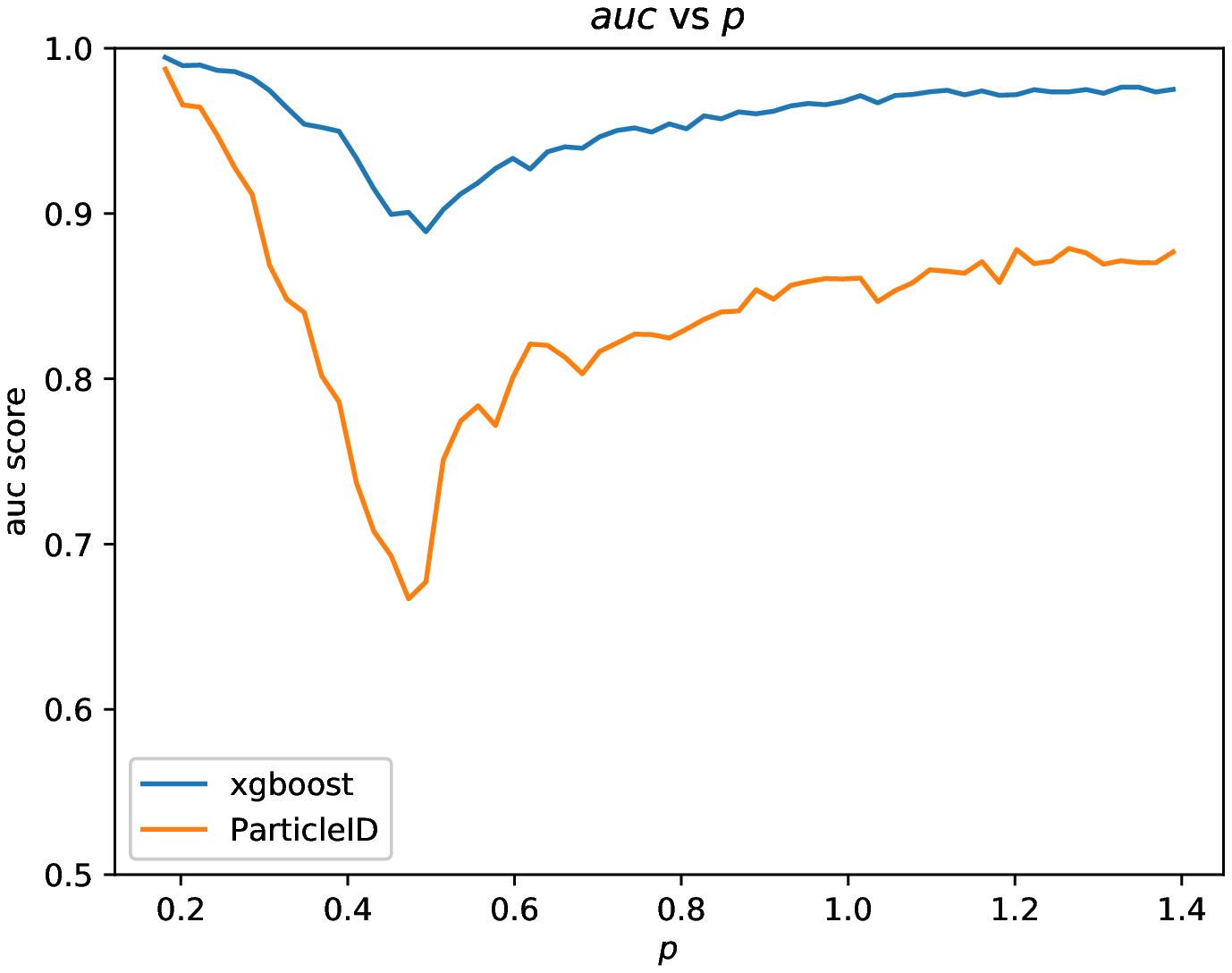}
\put(-175,60){{(a)}}
\put(-30,60){{(b)}}
\caption{Comparison between new approach with XGBoost (blue curve) and the default PID (yellow curve) (a) ROC curve (b) AUC curve.}
\label{pid_roc_auc}       
\end{figure}

We propose a nesting architecture with XGBoost~\cite{XGBoost} classifiers for $\mu$ identification. We use two classifiers with all the reconstructed information from EMC and MUC as inputs, respectively. The outputs of the two classifiers together with $\chi^{2}_{\mathrm{dE/dX}}$ and $\chi^{2}_{\mathrm{TOF}}$ are submitted to another classifier for combination.
Based on studies with MC samples, Figure~\ref{pid_roc_auc} shows the new approach with ML has a better performance than the default muon identification used in BESIII. The performance drops around 0.4~GeV because there is a cut-off of the MUC for those low momentum particles cannot reach the muon counter.

\section{Reweighting with XGBoost}
\begin{figure}[htbp]
\centering
\includegraphics[width=4cm]{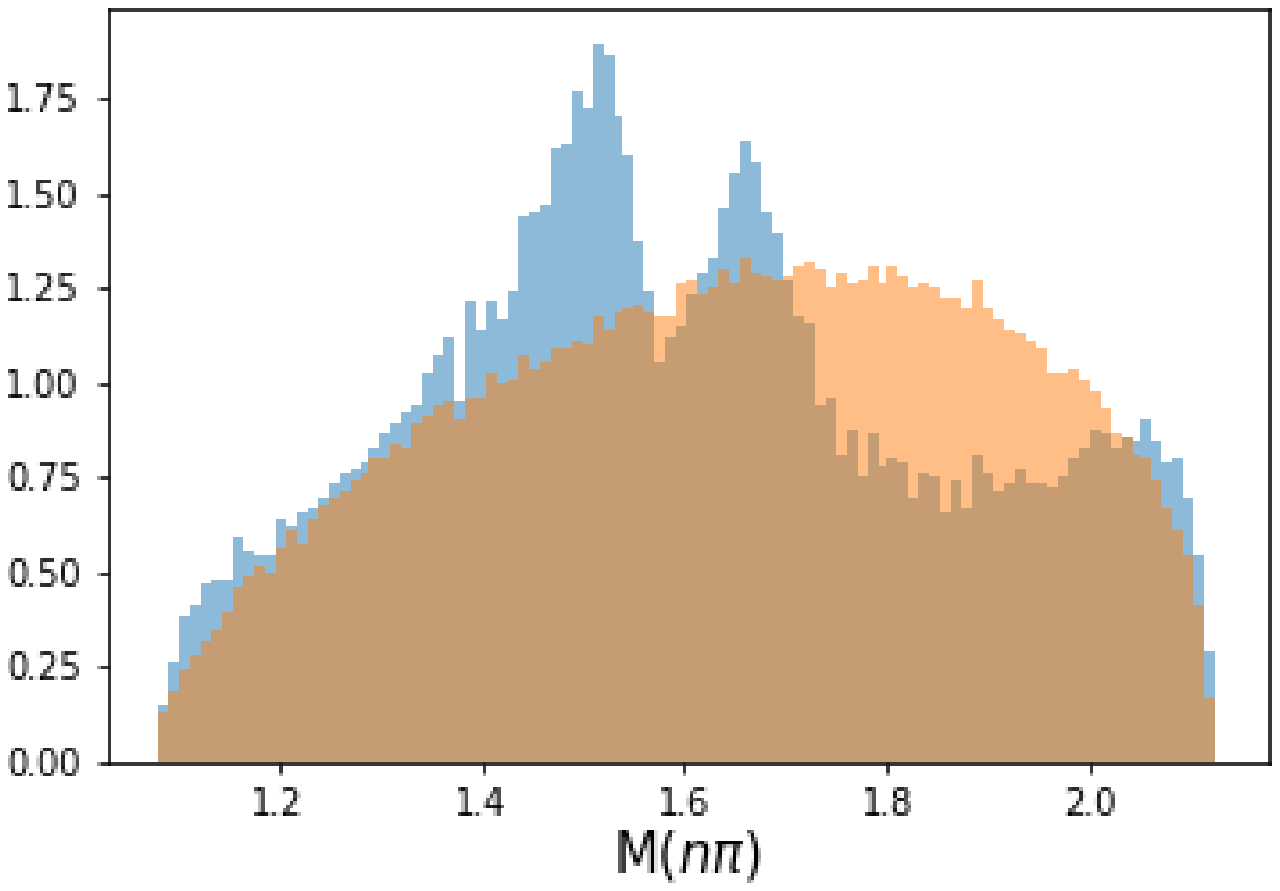}\put(-30,55){{(a)}}
\includegraphics[width=4cm]{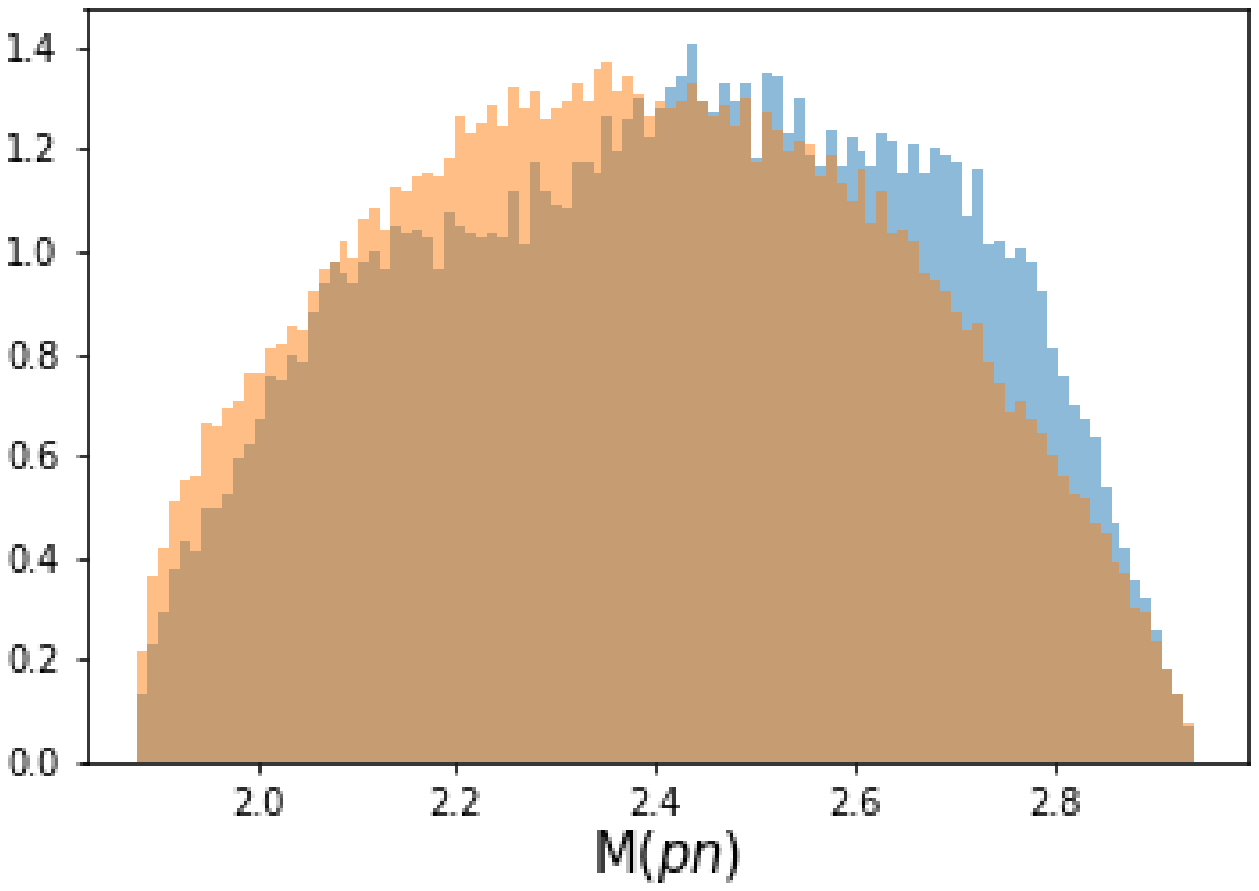}\put(-30,55){{(b)}}
\includegraphics[width=4cm]{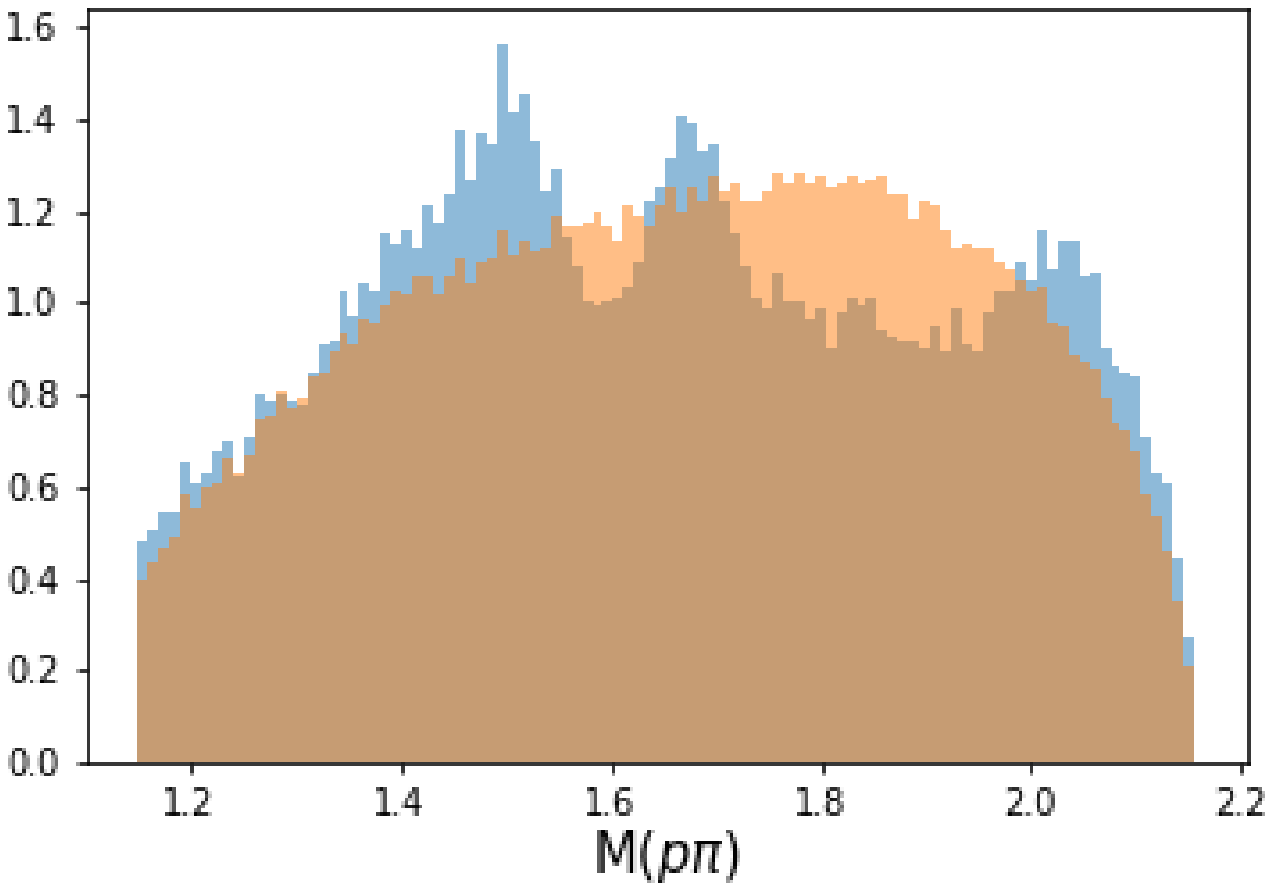}\put(-30,55){{(c)}}
\vskip -0.3cm
\includegraphics[width=4cm]{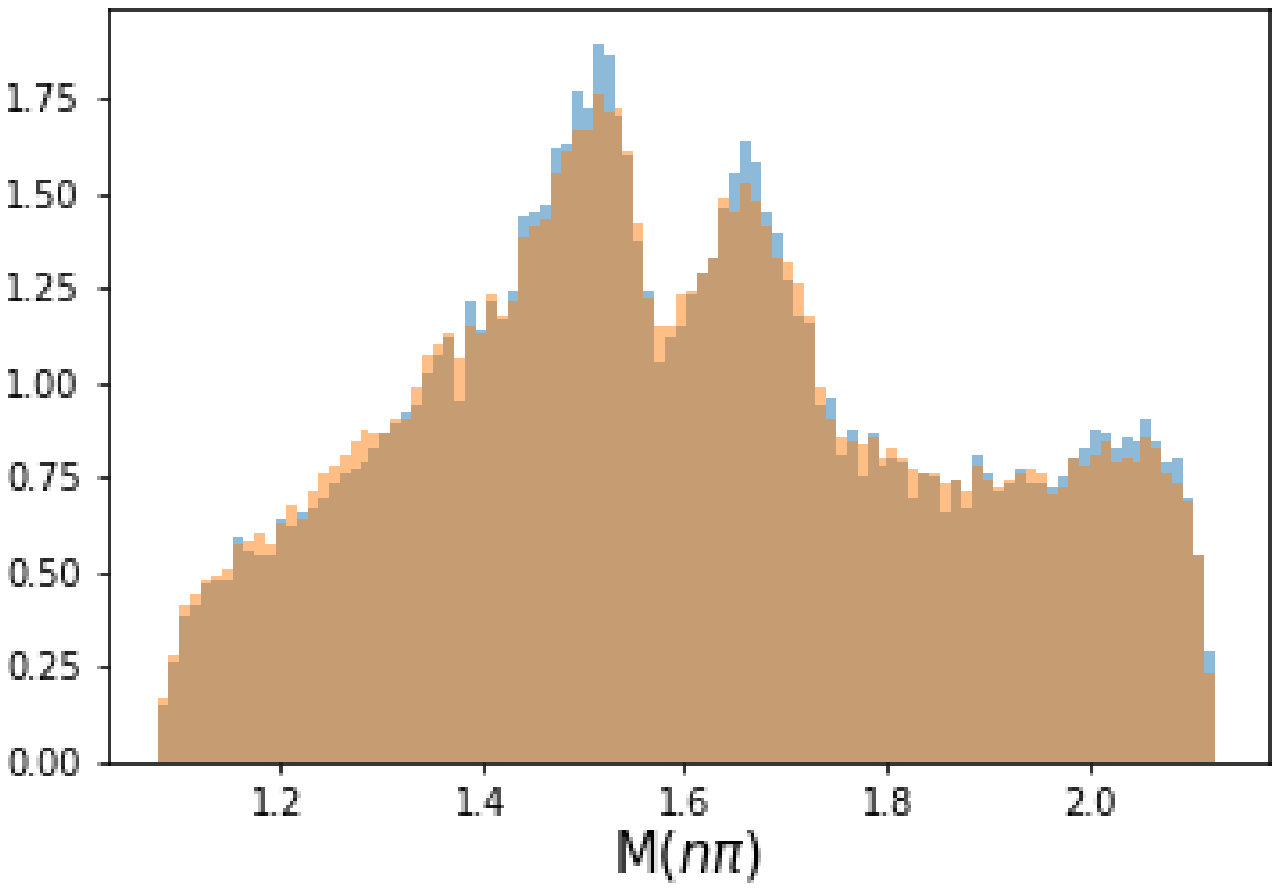}\put(-30,55){{(d)}}
\includegraphics[width=4cm]{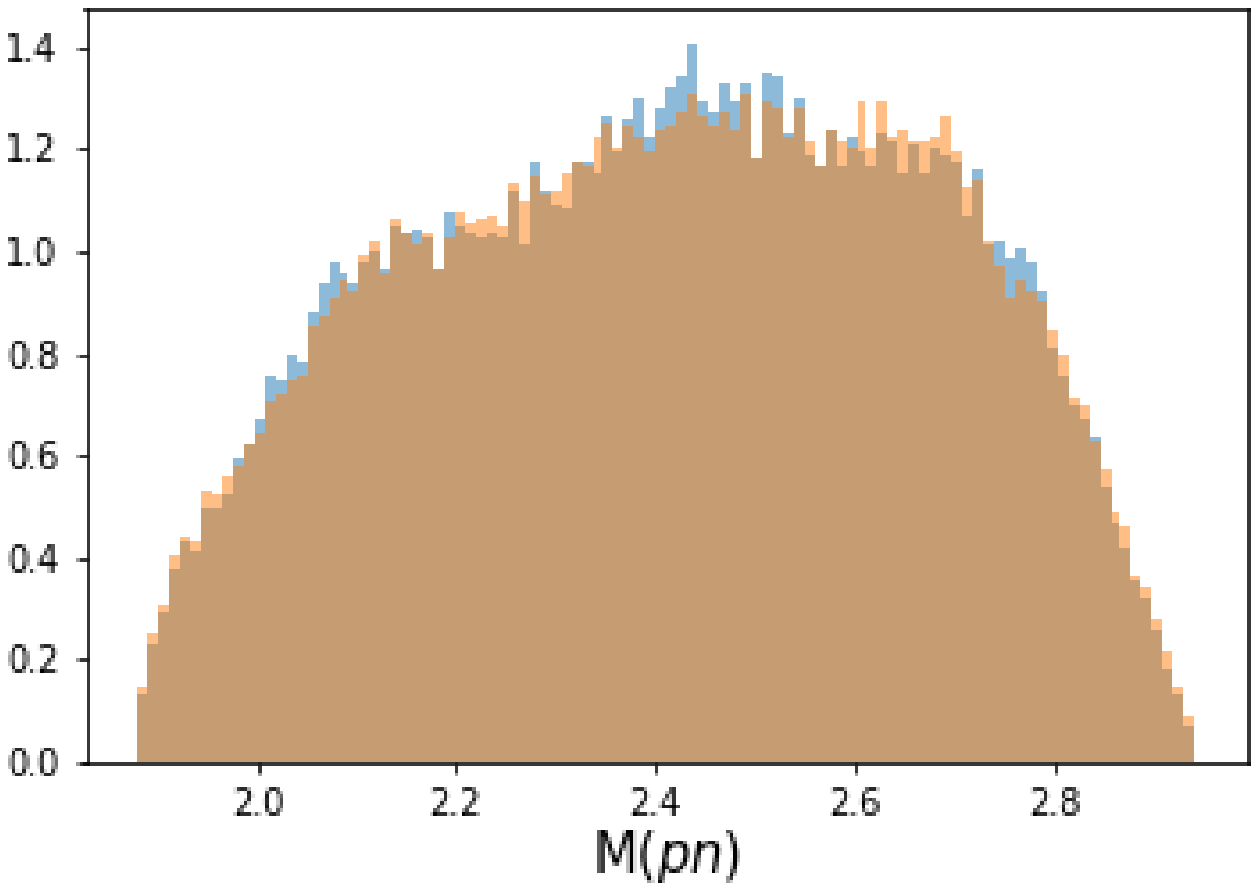}\put(-30,55){{(e)}}
\includegraphics[width=4cm]{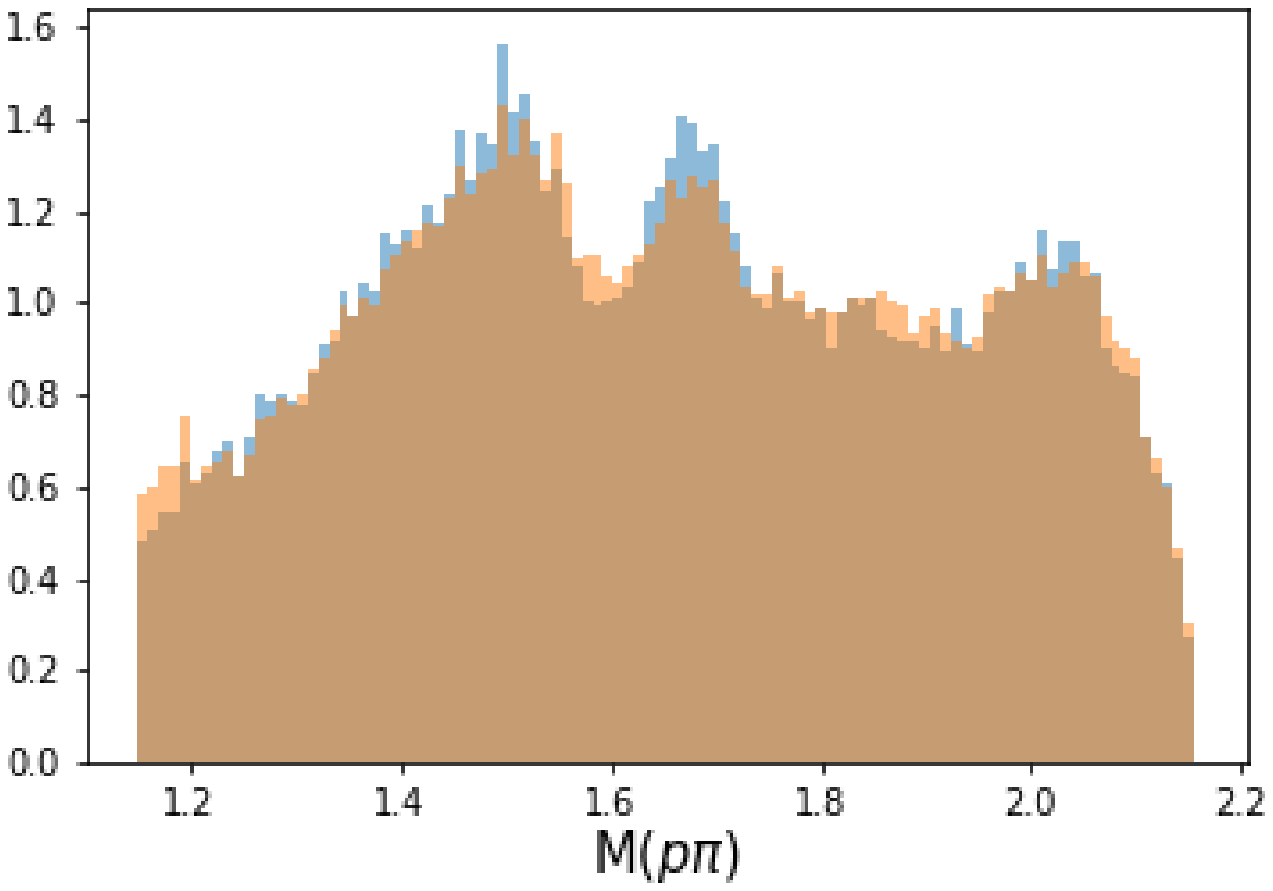}\put(-30,55){{(g)}}
\caption{Comparison of pseudo data (blue) and PHSP (yellow) distributions before(a-c) and after (d-f)
using the XGBoost reweighter.}

\label{dis_before}
\end{figure}

Modeling data with Monte Carlo (MC) simulation is essential for physics analysis, e.g., for calculation of efficiency and estimation of background. In the energy regime of non-perturbative QCD, experimental data of BESIII is rich of resonances, which cannot be fully described with a generic MC model. Amplitude analysis~\cite{pwa_besiii} is usually required to extract the properties of resonances and model the data.
In the case that results of amplitude analysis are not available yet, it is very useful to create ``data-like'' MC by multi-dimensional reweighting to correct for differences between data and MC.
There are two methods for reweighting with ML techniques ~\cite{ML_reweight1, ML_reweight2}  have been proposed. We utilize the approach~\cite{ML_reweight1} with XGBoost algorithm for physics analysis at BESIII~\cite{reweight_xxa}.
For illustration, a MC sample of $J/\psi\to N^* \bar{n} +c.c.\to p\pi^- \bar{n} +c.c.$ including a set of intermediate $p\pi$ resonances is generated as pseudo data. Reweighting is applied to a sample of phase-space-distributed MC (PHSP). Figure~\ref{dis_before} shows the results after reweighting.

\section{Cluster reconstruction of cylindrical GEM inner tracker}
\begin{figure}[h]
\centering
\includegraphics[width=7.cm,clip]{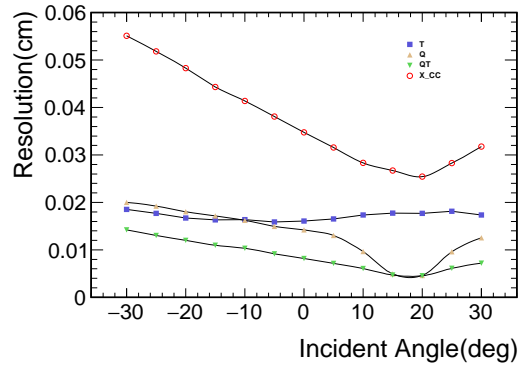}
\caption{Resolution curve along with incident angle, blue curve for T input only, yellow curve for Q input only, green curve for Q, T input together, red curve for charge centroid results. }
\label{cgem_reso}       
\end{figure}

Due to the aging of the inner drift chamber, an upgrade of inner tracker with 3 layers of cylindrical triple-GEMs is scheduled for BESIII in 2019.
As the input of track reconstruction, cluster reconstruction is to measure the position of the ionizing particle in the drift cathode layer with the readouts from the anode strips.
There are two methods for cluster reconstruction of CGEM-IT~\cite{cgem_recon}.
The charge centroid method calculates the weighted average position of the strips with their charge (Q). The method based on time measurements(T) is using the drift gap as a ``micro time projection chamber'' (micro-TPC)~\cite{Alexopoulos:2010zz}. To improve the position resolution, the results of the two methods can be further combined according to their resolutions and correlations. However, the correlations between resolution and incident angle are quite complicated and difficult to handle.
A ML method based on XGBoost regressor is proposed to reconstruct the initial ionizing particle position from the readouts (Q and T) of the fired strips. A simulation with a standalone digitization code, based on GARFIELD~\cite{garfield}, is used to generate the event with 1~T magnetic field, incident angle between -30$^\circ$ to 30$^\circ$ for one layer of planar Triple-GEM.
The results compared with the charge centroid method are shown in Figure~\ref{cgem_reso}. The resolution of the XGBoost regressor with Q or T inputs alone is better than that of the charge centroid method. The results with Q have larger dependency with incident angle. The combined result with Q and T by ML is further improved.
\section{Summary}
Three new use cases of ML at BESIII are presented in this proceeding and show promising results. In the future, we can further investigate the application of ML to improve the performance of BESIII software, e.g., the tracking of low momentum charged particles, the tracking with high background rates,~{\it etc.}


\begin{thebibliography}{}
%
%
\bibitem{phsysics_at_besiii}
D. M. Asner  ~\textit{et al.}, Int. J. Mod. Phys. A \textbf{24}, S1-794 (2009)
\bibitem{besiii_detector}
M. Ablikim ~\textit{et al.} (BESIII Collaboration), Nucl. Instrum. Methods Phys. Res., Sect. A \textbf{614}, 345 (2010)

\bibitem{XGBoost}
T. Chen , C. Guestrin,  Proceedings of the 22Nd ACM SIGKDD International Conference on Knowledge Discovery and Data Mining (785-794) (2016)
\bibitem{pwa_besiii}
M. Ablikim ~\textit{et al.} (BESIII Collaboration), Phys. Rev. D \textbf{88} 112007, (2013)
\bibitem{ML_reweight1}
D. Martschei  ~\textit{et al.}, Journal of Physics: Conference Series, 368(1), 012028 (2012)
\bibitem{ML_reweight2}
A. Rogozhnikov,  Journal of Physics: Conference Series, 762(1), 012036 (2016)
\bibitem{reweight_xxa}
Xian Xiong, $\&$ Beijiang Liu. (2018, October 9).
Zenodo. http://doi.org/10.5281/zenodo.1451985
\bibitem{cgem_recon}
R. Farinelli  ~\textit{et al.}, arXiv:1807.00500 (2018)
\bibitem{Alexopoulos:2010zz}
  T.~Alexopoulos {\it et al.},
  Nucl.\ Instrum.\ Meth.\ A {\bf 617}, 161 (2010).
\bibitem{garfield}
R.Farinelli, L.Lavezzi ~\textit{et al.}, arXiv:1807.01210 (2018)

\end{thebibliography}
\end{document}